\documentclass[twoside,slac_one]{revtex4}
\usepackage{graphicx}
\usepackage{fancyhdr}
\usepackage{amsmath} 
\usepackage{bm}
\usepackage{amsxtra}
\usepackage{amssymb}
\usepackage{amsthm}
\usepackage{latexsym}
\usepackage{lscape}

\newcommand{\met}{\rm\,/\!\!\!\!{\it E_T}}
\newcommand{\tn}[1]{\textnormal{#1}}

\begin{document}

\title{Search for Chargino-Neutralino Associated Production in Dilepton Final States with Tau Leptons}

%

\author{R. Forrest}
\affiliation{Department of Physics, UC Davis, Davis, CA, USA}
\author{M. Chertok}
\affiliation{Department of Physics, UC Davis, Davis, CA, USA}

\author{(on behalf of the CDF Collaboration)}

\begin{abstract}
We present a search for chargino and neutralino supersymmetric particles yielding same signed dilepton final states including one hadronically decaying tau lepton using 6.0 $fb^{-1}$ of data collected by the the CDF II detector.  This signature is important in SUSY models where, at high $\tan{\beta}$, the branching ratio of charginos and neutralinos to tau leptons becomes dominant. We study event acceptance, lepton identification cuts, and efficiencies. We set limits on the production cross section as a function of SUSY particle mass for certain generic models. 
\end{abstract}

\maketitle

\thispagestyle{fancy}


\section{Introduction}
In the search for new phenomena, one well-motivated extension to the Standard Model (SM) is supersymmetry (SUSY).  One very promising mode for SUSY discovery at hadron colliders is that of
chargino-neutralino associated production with decay into three leptons. Charginos decay into a
single lepton through a slepton $$\tilde{\chi}_1^{\pm}  \rightarrow ~ \tilde{l}^{(*)} ~\nu_l \rightarrow
\tilde{\chi}_1^{0} ~l^{\pm} ~\nu_l $$ and neutralinos similarly decay into two detectable
leptons $$\tilde{\chi}_2^{0}  \rightarrow ~ \tilde{l}^{\pm(*)} ~l^{\mp}
\rightarrow \tilde{\chi}_1^{0} ~l^{\pm} ~l^{\mp} $$. 
The detector signature is thus three SM leptons with associated missing energy from the undetected neutrinos and lightest neutralinos, $\tilde{\chi}_1^0$ (LSP), in the event. Many previous searches have used all three leptons for detection \cite{rut_note,2009arXiv0910.1931F}. 

The most generic form of SUSY is the MSSM model which, in many parameter spaces, gives the lepton signature that interests us \cite{susy_primer}. Unfortunately there are far too many free parameters in this model to test generically.  In the past it has been tradition to use a specific gravity mediated SUSY breaking model called mSugra.  For this analysis we adopt a more generic method, in which we present results in terms of exclusions in sparticle masses as opposed to mSugra parameter space.

We construct simplified models of SUSY wherein we do not hope to develop a full model of SUSY, but an effective theory that can be easily translated to describe kinematics of arbitrary models. We set the masses at the electroweak scale and include the minimal suite of particles necessary to describe the model and effectively decouple all other particles, by setting their masses $> $ TeV range. We also tune the couplings of the particles to mimic models that preferentially decay to taus.

Specific models will determine permitted decay modes \cite{Ruderman:2010kj}.  Different models'  SUSY breaking method will determine allowed decay modes in broad categories.  In this analysis we present two types of generic models. The first is a simplified gravity breaking model similar to mSugra; the second is a simplified gauge model, which encompasses a broad suite of theories with gauge mediated SUSY breaking (GMSB).

The simplified gravity model we generally have electroweak ($W^{\pm}$) production of $\tilde{\chi}_1^{\pm}, \tilde{\chi}_2^{0}$ pairs. $\tilde{\chi}_1^{\pm}$ then decays to $\tilde{l}^{\pm}, \nu_l$ and $\tilde{\chi}_2^{0}$ goes to $\tilde{l}^{\pm}  l^{\mp}$. All the sleptons decay as normal $\tilde{l}^{\pm} \rightarrow l^{\pm}, \tilde{\chi}_1^{0}$. We can tune the branching ratio to slepton flavors. For each SUSY point, we choose two branching ratios BR($\tilde{\chi}_2^{0}, \tilde{\chi}_1^{\pm} \rightarrow \tilde{\tau} + X) = 1, 1/3$. We choose the masses of the $\tilde{\chi}_1^{\pm}$ and $\tilde{\chi}_2^{0}$ to be equal. 

The simplified gauge model is motivated by gauge mediated SUSY breaking scenarios. Generally, the LSP is the gravitino which is very light: in the sub-keV range.  Also,  charginos do not couple to right handed sleptons in these models, therefore all chargino decays are to taus, so BR($\tilde{\chi}_1^{\pm} \rightarrow \tilde{\tau}_1 \nu_{\tau}) = 1 $ always. The  $\tilde{\chi}_2^{0}$ can decay to all lepton flavors. The final feature of this model is that $\tilde{\chi}_1^{\pm}$ or $\tilde{\chi}_2^{0}$  don't decay through SM bosons. 

\section{Analysis Overview}
\label{sec:overview}
Our approach is to look for the two same signed leptons from trilepton events since the opposite signed pair has the disadvantage of large standard model backgrounds from electroweak Z decay. 

We select one electron or muon and one hadronically decaying tau. Requiring a hadronicaly decaying tau adds sensitivity to high tan$ \beta$ SUSY space. Our main backgrounds therefore will be SM W + Jets where the W boson decays to an electron or muon and the jet fakes a hadronic tau in our detector.

Our background model is comprised of two distinct types. We use Monte Carlo to account for common SM processes naturally entering the background as well as processes with real taus that might contain a fake lepton. Any process involving a jet faking a tau is covered in our tau fake rate method, these processes would be W + Jet, conversion+Jet and QCD. In all these processes, the jet fakes a tau and a lepton comes from the other leg of the event. 

Our fake rate is measured in a sample of pure QCD jets \cite{2009PhRvL.103t1801A}. We validate the measurement by applying it to three distinct orthogonal regions to our signal.

We select our dilepton events and first understand the opposite signed lepton-tau region. After applying an $H_T$ cut, we develop confidence that we understand the primary and secondary backgrounds, $Z\rightarrow \tau \tau$ and W + jets respectively. We then look at the same signed signal region, where we expect to be dominated by our fake rate background.

To set limits in the M(Chargino) vs. M(Slepton) plane, a grid of signal points is generated. We optimize a $\met$ cut as a function of model parameters for each point to increase our sensitivity to signal. Limits are then found at each point, and iso-contours are interpolated to form our final limits on SUSY process cross section.

\section{Dataset And Selection}
\label{sec:data}
We use 1.96 TeV $p\bar{p}$ collision data from the Fermilab Tevatron corresponding to 6.0 $fb^{-1}$ of integrated luminosity from the CDF II detector. The data is triggered by requiring one lepton object, and electron or muon; as well as a cone isolated tau like object. We then apply standard CDF selection cuts to the objects. Electrons and muons are required to have an $E_T$ ($P_T$) cut of 10 GeV. One pronged taus have a $P_t$ cut of 15 GeV/c and three pronged taus have a 20 GeV/c cut. The $P_T$ for a tau is considered to be the visible momentum: the sum of the tracks and $\pi^0$'s in the isolation cone.

To reduce considerable QCD backgrounds we apply a cut on $H_T$ defined as the sum of the tau, lepton and $\met$ in the event. The $H_T$ cut is 45,50,55 GeV/c for the $\tau_1-\mu$, $\tau_1-e$ and $\tau_3 - \ell$ channels. We cut events were $d\phi (l,\tau) < 0.5$ as well as events with OS leptons within 10 GeV of the Z boson mass. $\met$ is corrected for all selected objects and any jets observed in the event.

Monte Carlo is scaled to reflect trigger inefficiencies as well as inefficiencies from lepton and tau reconstruction. 

\section{Backgrounds}
Our background model is comprised of two distinct types. We use Monte Carlo to simulate detector response to Diboson, $t\bar{t}$, Z boson processes as well as real taus from W decay. These processes are normalized to their SM cross section and weighted by scale factors to account for inefficiencies in trigger, ID and reconstruction. Any process involving a jet faking a tau is covered in our tau fake rate method, these processes would be W + Jet, conversion+Jet and QCD. In all these processes, the jet fakes a tau and a lepton comes from the other leg of the event. 

We measure the fake rate in a sample of QCD jets. Our rate is defined as the ratio of tau objects to loose taus where loose taus are tau like objects that pass our trigger. Because the trigger has very decent tau discriminating ability, this relative fake rate is fairly high. In terms of applying the fake rate to fakeable objects, in order to not overestimate our fake contributions we have a subtraction procedure for the preponderance of real taus that pass through our trigger. The measurement of the fake rate in the leading jet and sub leading QCD jet constitutes the systematic on the measurement.

We validate our tau fake rates in three different orthogonal regions to our signal. These regions reflect the three processes the fake rate will account for in the analysis.

\section{OS Validation}
\begin{figure}[h!]
  \begin{tabular}{|c|c|}
  \hline
    \includegraphics[width=8cm,clip=]{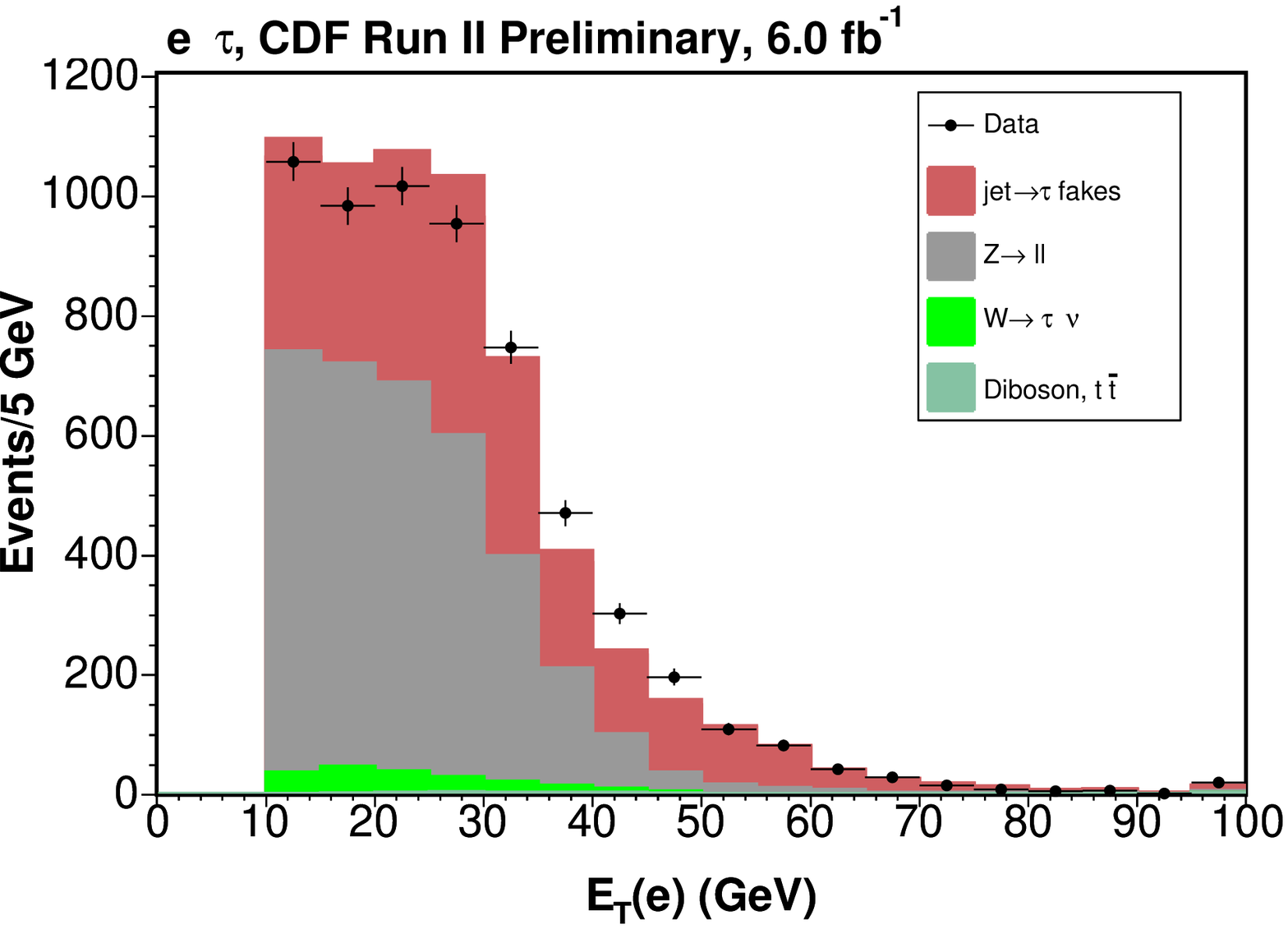} &   \includegraphics[width=8cm,clip=]{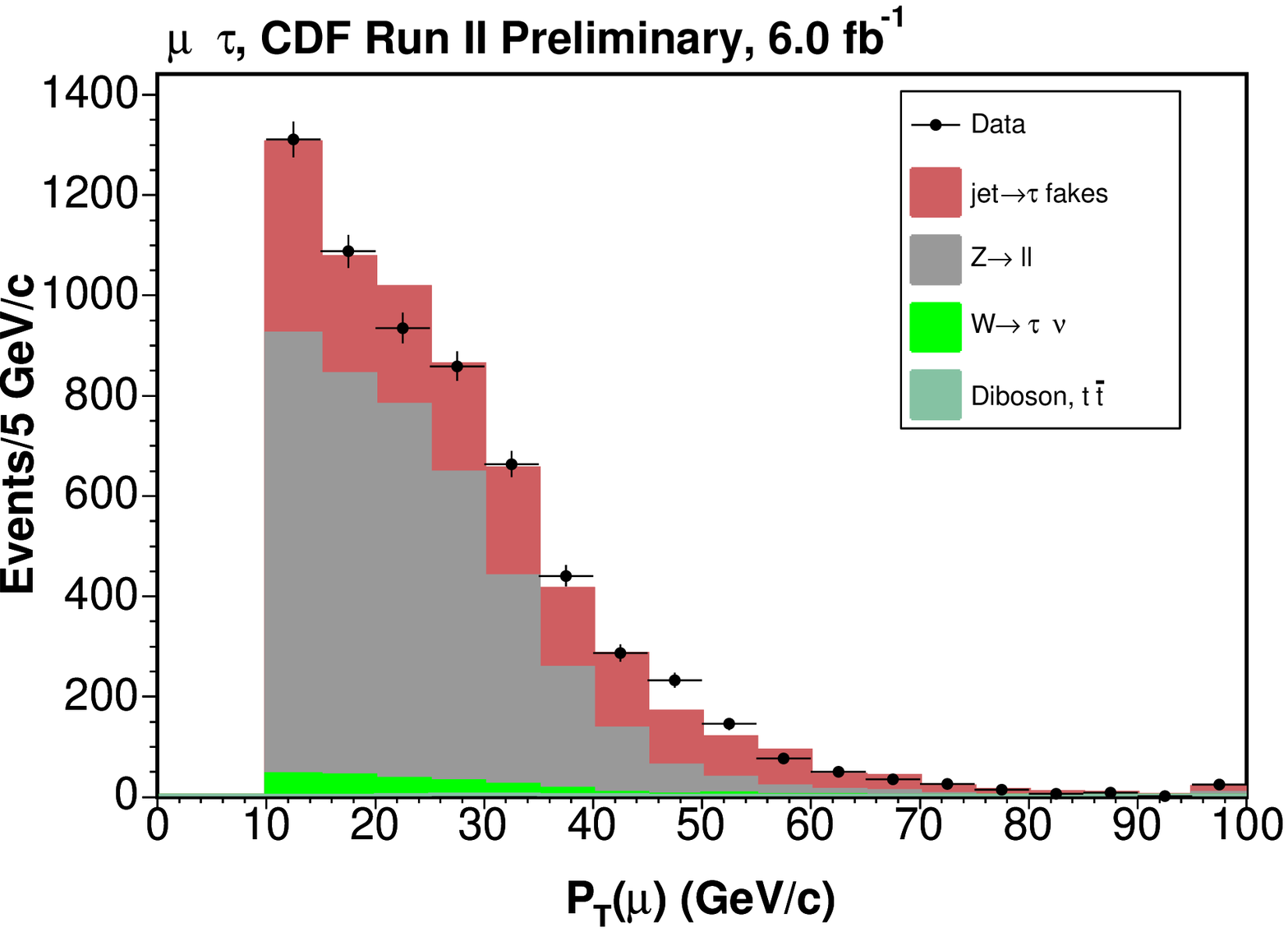}\\ 
\hline
  \end{tabular}
    \caption{Plots of the OS Control Region, Electron $E_T$ (left) and Muon $P_T$ (right).  \label{fig:os_plots}}
\end{figure}

Before we look at signal data in out blind analysis, a major validation step is to confirm agreement in the OS region. This region is dominated by $Z\rightarrow \tau \tau$ decays, which gives us confidence in our scale factor application. The secondary background in this region is W+ Jets, which serves as an additional check on our fake rate background. As can be seen in Table~\ref{table:oscr} as well as in Figure~ \ref{fig:os_plots} and we have good confidence in our background model.

\begin{table}[h]
 \centering
 \mbox{
 \begin{tabular}{|l|r|}
\hline \multicolumn{2}{|l|}{CDF Run II Preliminary $6.0\ \textrm{fb}^{-1}$} \\
\multicolumn{2}{|l|}{OS \  $ \ell - \tau $} \\ \hline
\hline Process & Events  $\pm$ stat $\pm$ syst  \\ \hline
Z$\rightarrow \tau \tau $ & $6967.3\pm 56.4\pm 557.4$  \\
Jet$\rightarrow \tau $ & $4526.5  \pm 26.8 \pm 1064.5$  \\
Z$\rightarrow \mu \mu $ & $262.5 \pm  20.1 \pm 21.0$  \\
Z$\rightarrow e e $ & $82.5  \pm  8.6 \pm 6.6 $  \\
W$ \rightarrow \tau \nu $ &  $371.5  \pm 12.4 \pm 36.4 $  \\
t$\bar{\textrm{t}} $ & $36.3 \pm 0.3 \pm 5.1 $  \\
Diboson & $61.3 \pm 0.9 \pm 6.0 $  \\ \hline
Total & 12308.0 $\pm\ 67.3\pm 1202.3 $\\
Data & 12268\\ \hline
\end{tabular} }
  \caption{Total OS control region.} \label{table:oscr}
\end{table}

\subsection{Observed Data and Limit Setting}
After gaining confidence in the OS control region, we unblind the analysis and set limits on our models. For each signal point, we choose a $\met$ cut that optimizes the $s/\sqrt{b}$ at that point. To allow simple interpretation, we form an analytical expression for the $\met$ cut as a function of model parameters. Because of large QCD and conversion backgrounds at low $\met$ all limit setting is done above $\met = 20\ GeV$. The results are below in table~\ref{table:result_total_metcut}. Kinematic plots of the SS region are in Figure~\ref{fig:ss_plots}.

\begin{table}[h] 
 \centering 
 \mbox{ 
 \begin{tabular}{|l|r|} 
\hline \multicolumn{2}{|l|}{CDF Run II Preliminary $6.0\ \tn{fb}^{-1}$} \\ 
\multicolumn{2}{|l|}{SS \ $ \ell - \tau $} \\ \hline 
\hline Process & Events  $\pm$ stat $\pm$ syst  \\ \hline 
Z$\rightarrow \tau \tau $ & $10.2\pm 2.2\pm 0.8$  \\
Jet$\rightarrow \tau $ & $1152.7  \pm 15.2 \pm 283.1$  \\
Z$\rightarrow \mu \mu $ & $0.0 \pm  0.0 \pm 0.0$  \\ 
Z$\rightarrow e e $ & $0.0  \pm  0.0 \pm 0.0 $  \\
W$ \rightarrow \tau \nu $ &  $96.9  \pm 6.4 \pm 9.5 $  \\
t$\bar{\tn{t}} $ & $0.7 \pm 0.0 \pm 0.1 $  \\
Diboson & $4.3 \pm 0.2 \pm 0.4 $  \\ \hline
Total & 1264.8 $\pm\ 16.6\pm 283.3 $\\
Data & 1116\\ \hline
\end{tabular} } 
\caption{SS signal region used in limit setting,  $\met > 20 \ GeV $. Both Electron and Muon Channels.}
 \label{table:result_total_metcut}
  \end{table}
  
\begin{figure}[h!]
  \begin{tabular}{|c|c|}
  \hline
    \includegraphics[width=8cm,clip=]{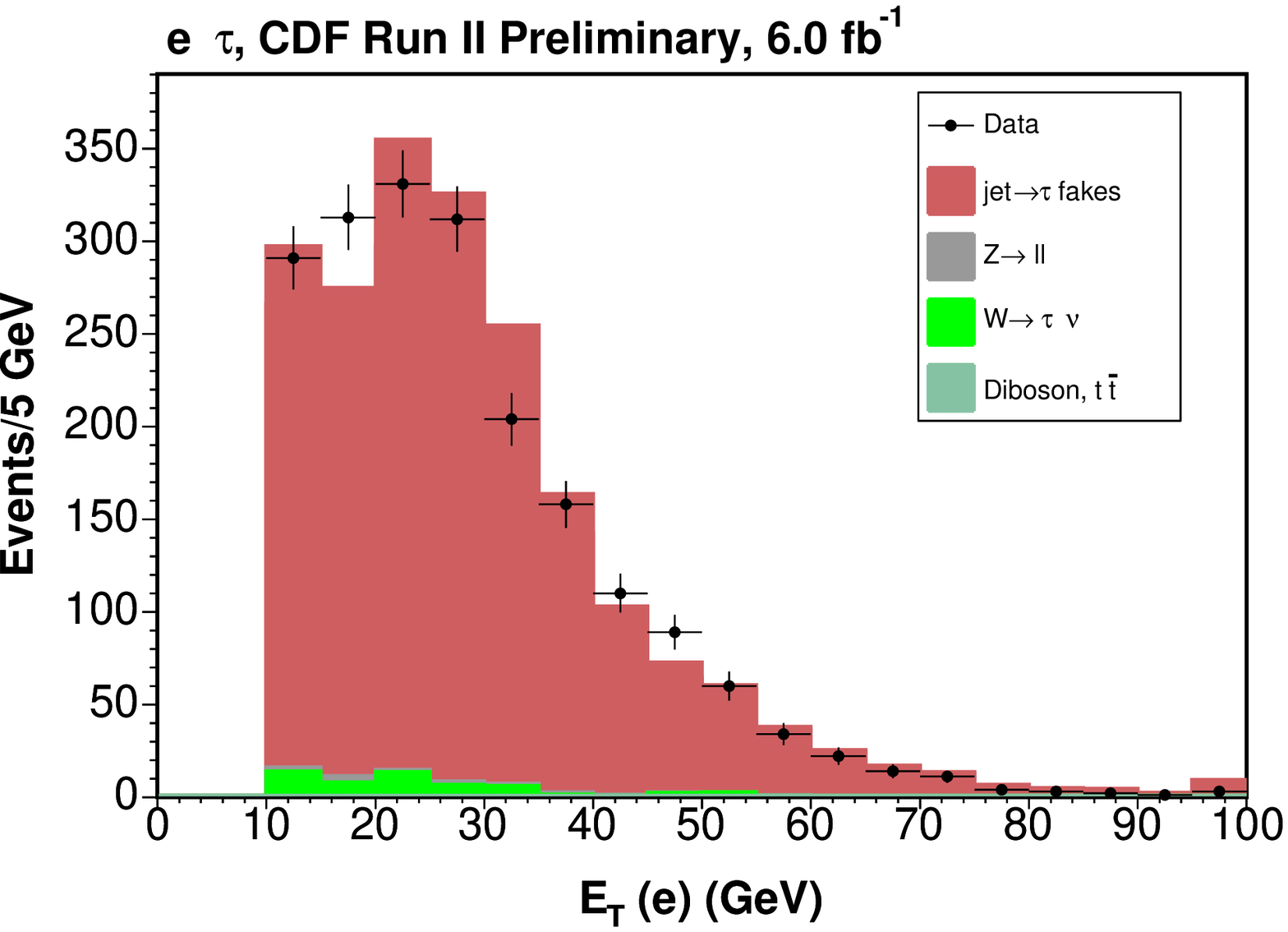} &   \includegraphics[width=8cm,clip=]{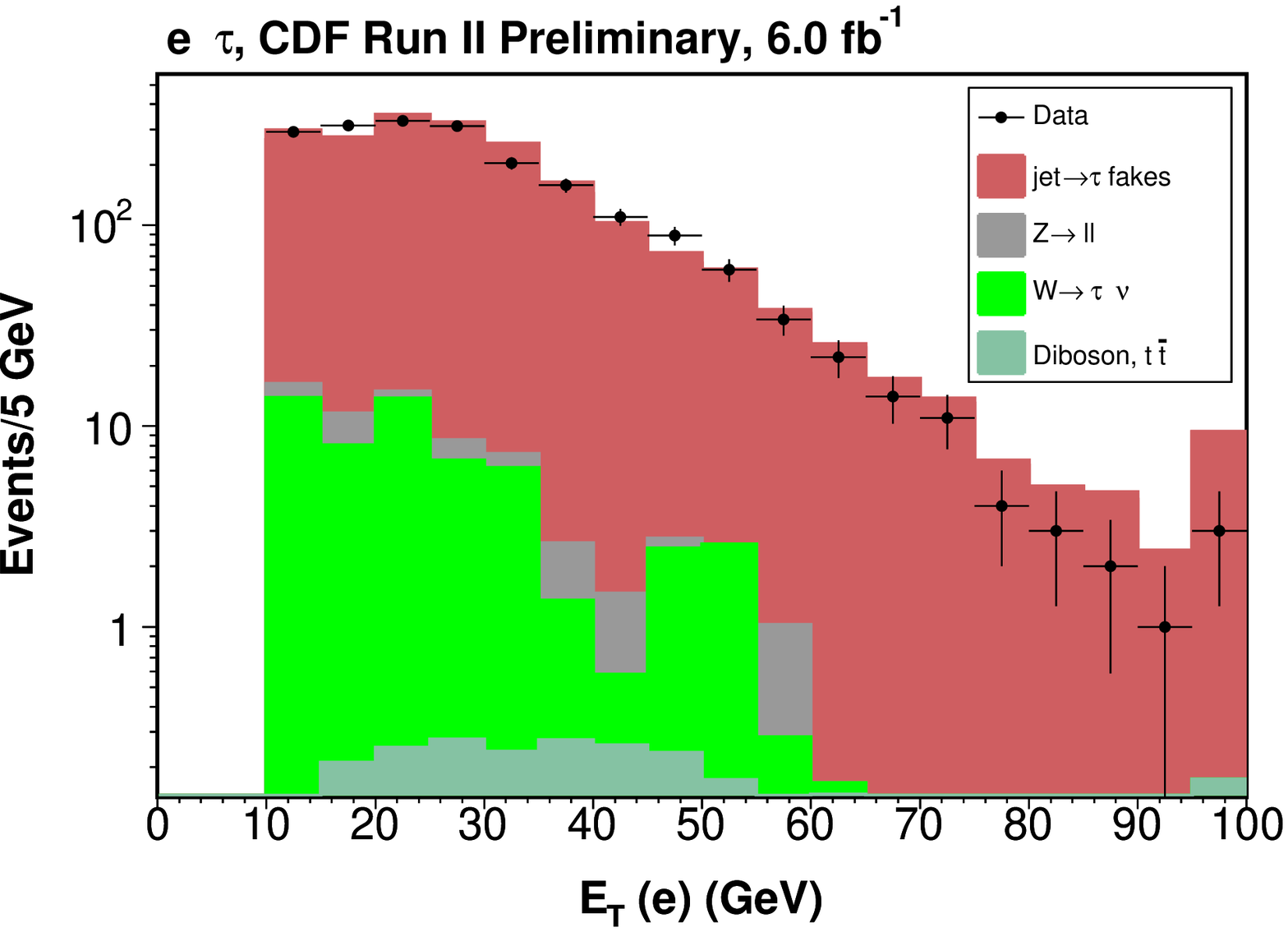}\\ 
\hline
  \end{tabular}
    \caption{Plots of the SS Signal Region, Electron $E_t$ (left) and a log version (right).  \label{fig:ss_plots}}
\end{figure}
\begin{figure}[h!]
  \begin{tabular}{|c|c|}
  \hline
    \includegraphics[width=8cm,clip=]{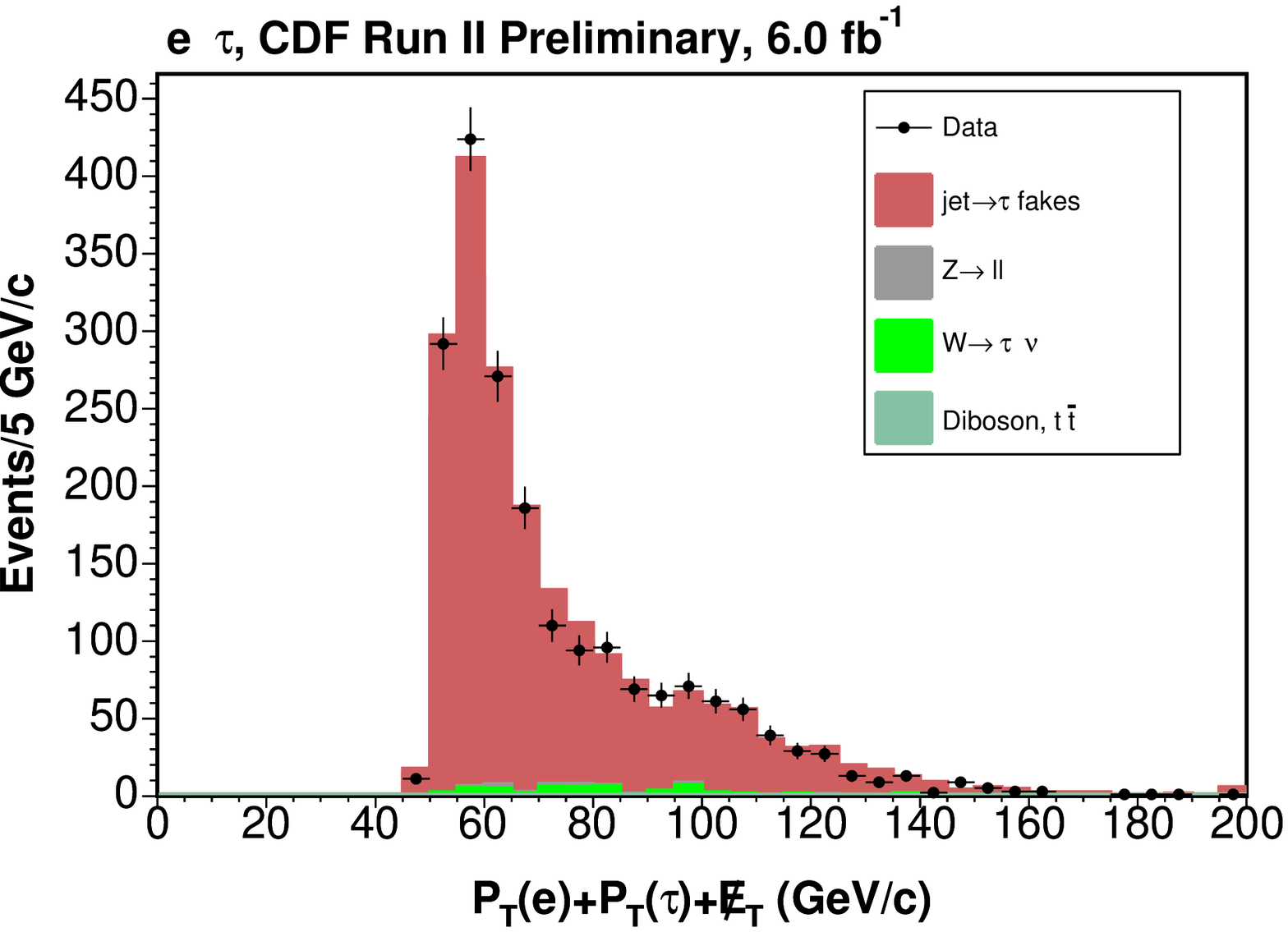} &   \includegraphics[width=8cm,clip=]{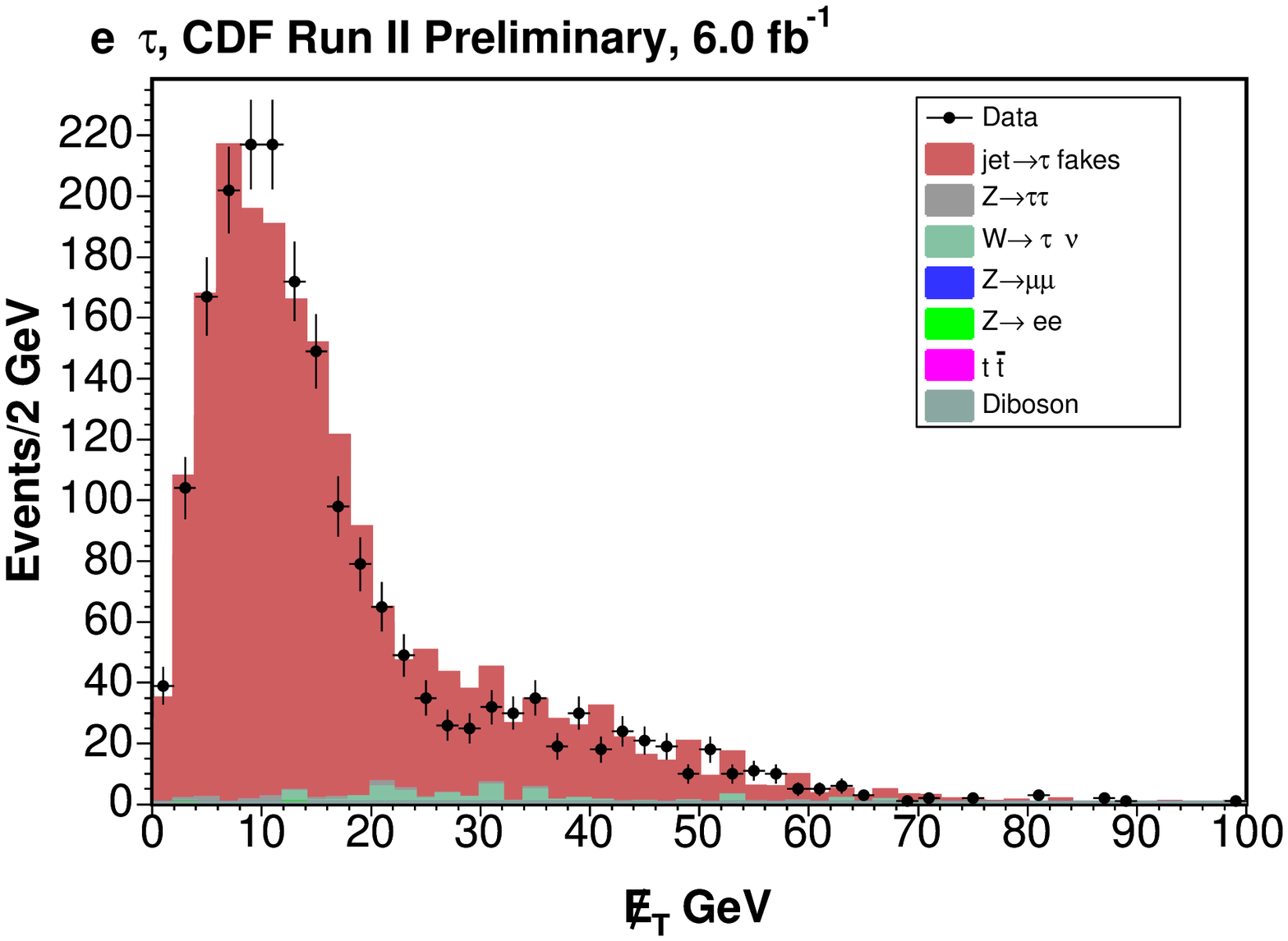}\\ 
\hline
  \end{tabular}
    \caption{Plots of the SS Signal Region, Electron $H_t$ (left) and a electron $\met$ (right).  \label{fig:ss_plots}}
\end{figure}

\begin{figure}[h!]
  \begin{tabular}{|c|c|}
  \hline
    \includegraphics[width=8cm,clip=]{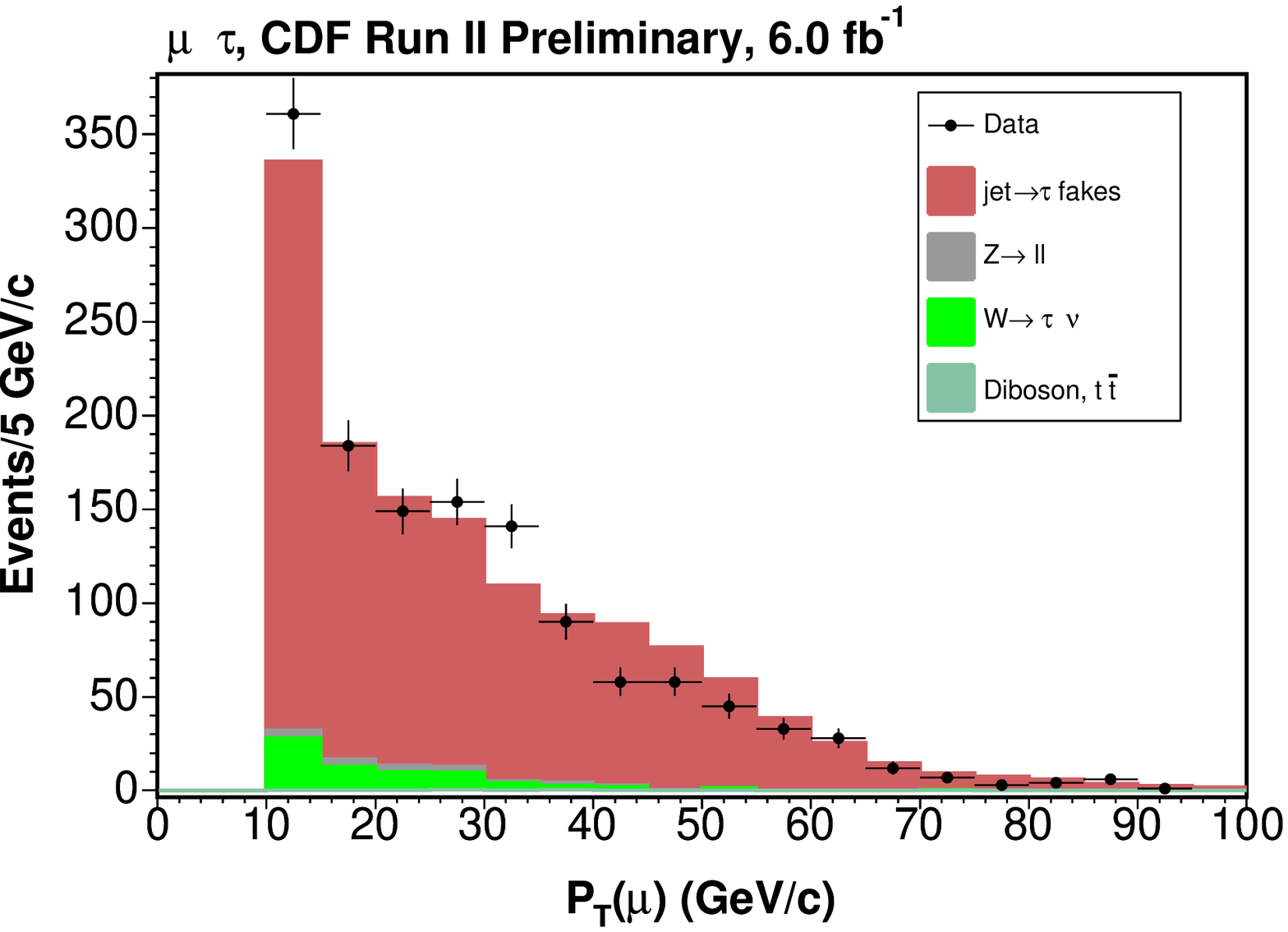} &   \includegraphics[width=8cm,clip=]{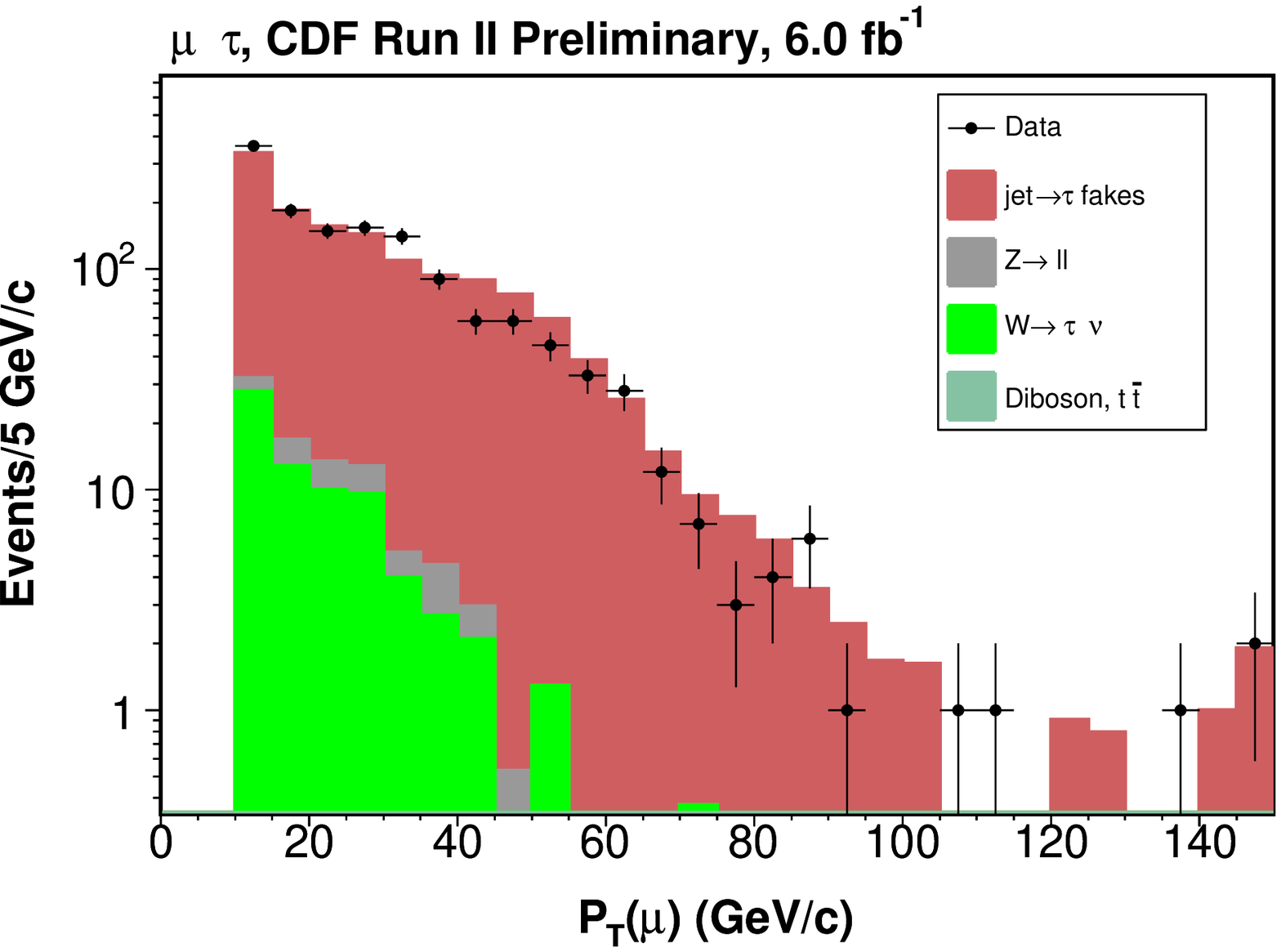}\\ 
\hline
  \end{tabular}
    \caption{Plots of the SS Signal Region, Muon $P_t$ (left) and a log version (right).  \label{fig:ss_plots}}
\end{figure}
\begin{figure}[h!]
  \begin{tabular}{|c|c|}
  \hline
    \includegraphics[width=8cm,clip=]{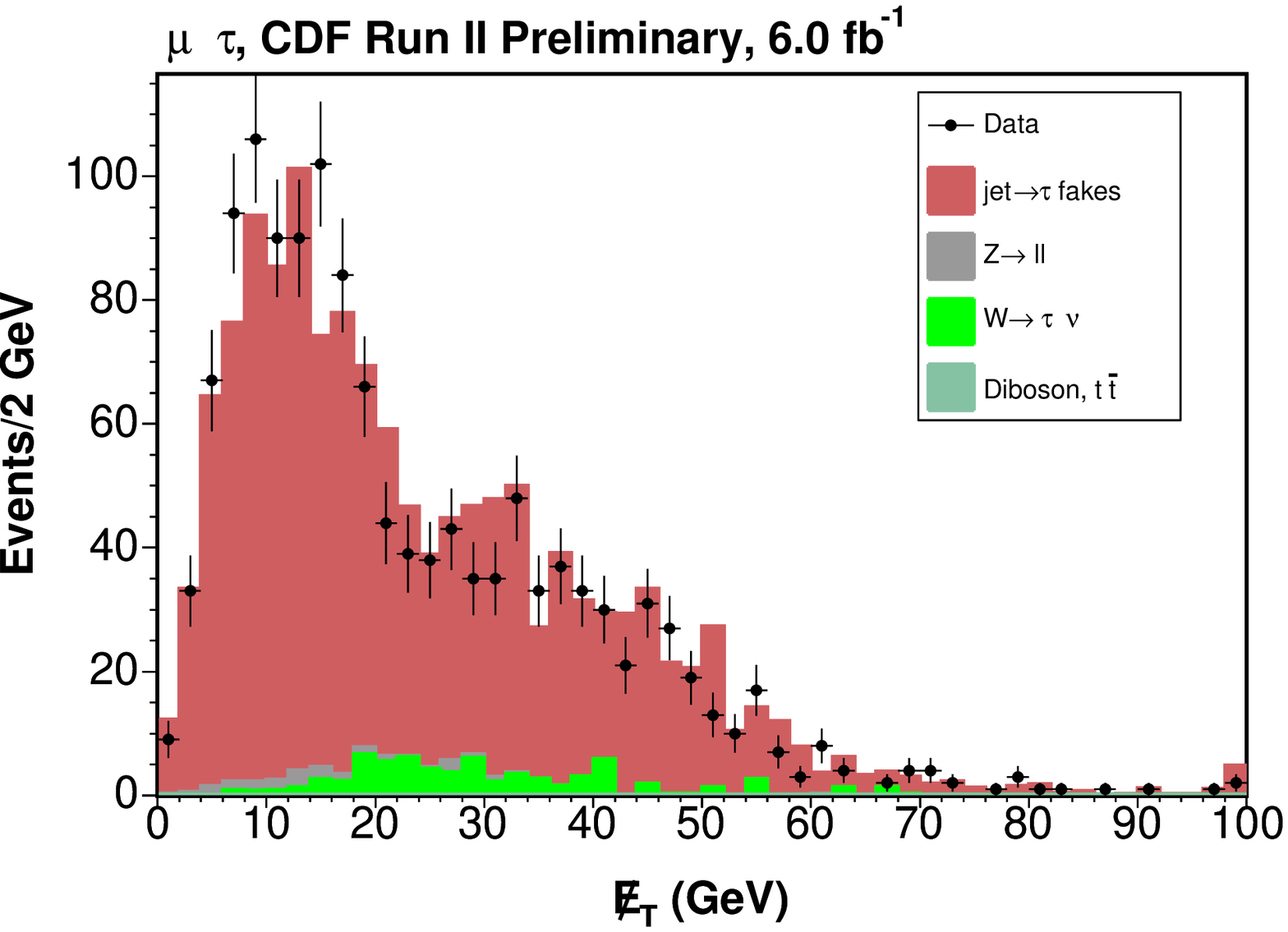} &   \includegraphics[width=8cm,clip=]{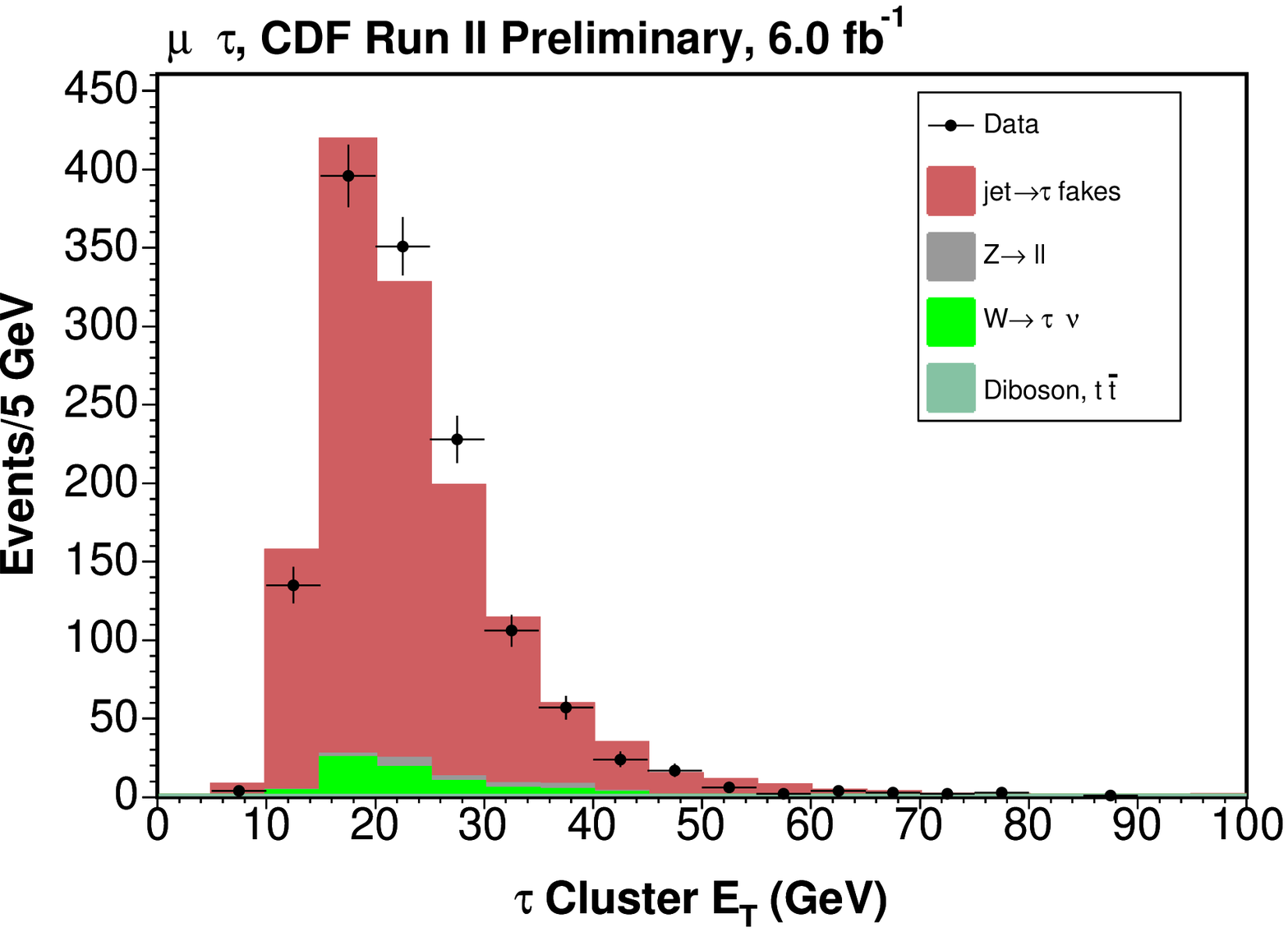}\\ 
\hline
  \end{tabular}
    \caption{Plots of the SS Signal Region, Muon $\met$ (left) and tau cluster $E_T$ (right).  \label{fig:ss_plots}}
\end{figure}

\begin{figure}[h!]
  \begin{tabular}{|c|c|}
  \hline
    \includegraphics[width=8cm,clip=]{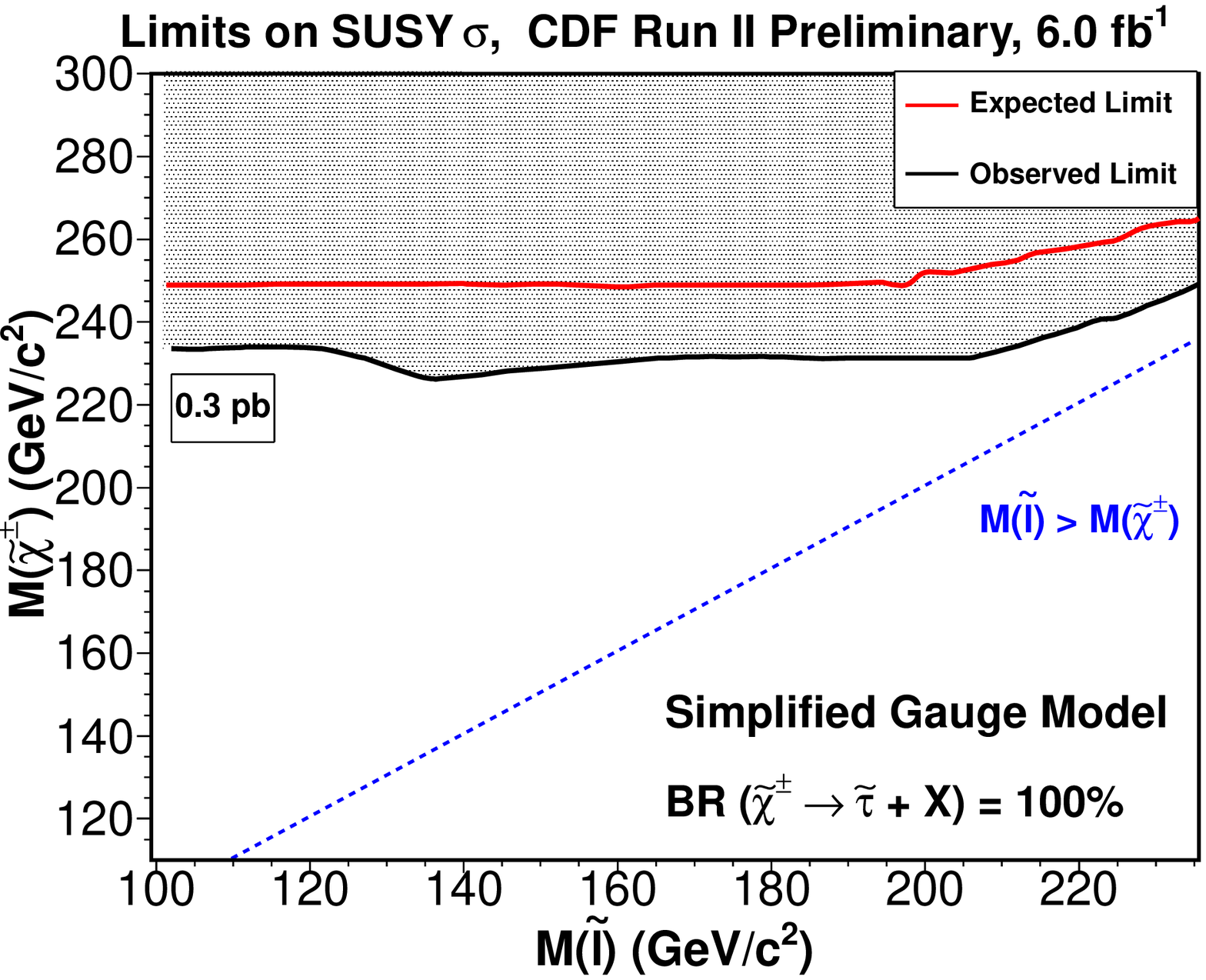}  &    \includegraphics[width=8cm,clip=]{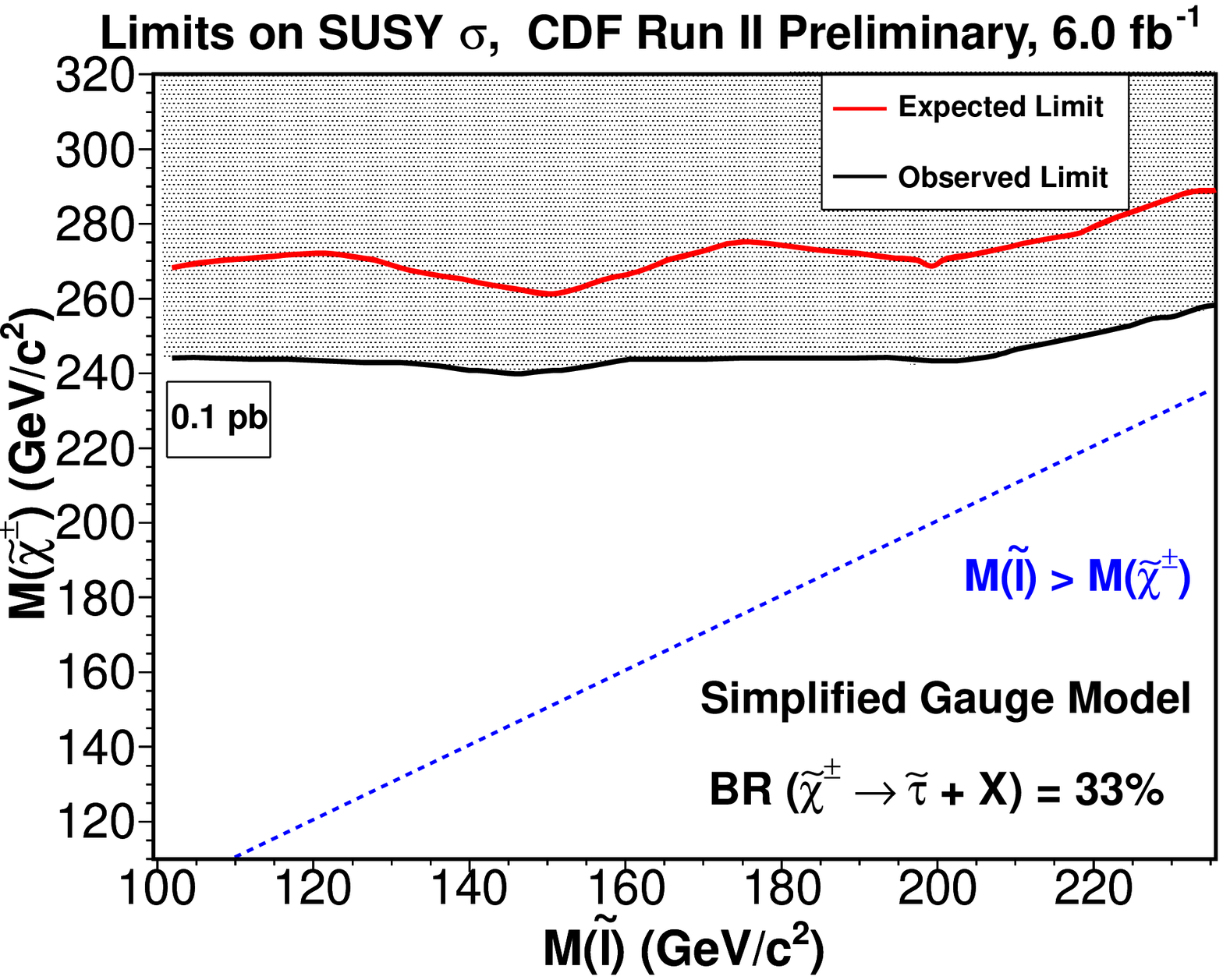} \\
\hline
  \end{tabular}
    \caption{Expected limits (pb) for Simplified Gauge Model for BR to taus of 100\% ( left),  and 33\%(right)  \label{fig:exp_gauge_2d}}
\end{figure}

\begin{figure}[h!]
  \begin{tabular}{cc}
    \includegraphics[width=8cm,clip=]{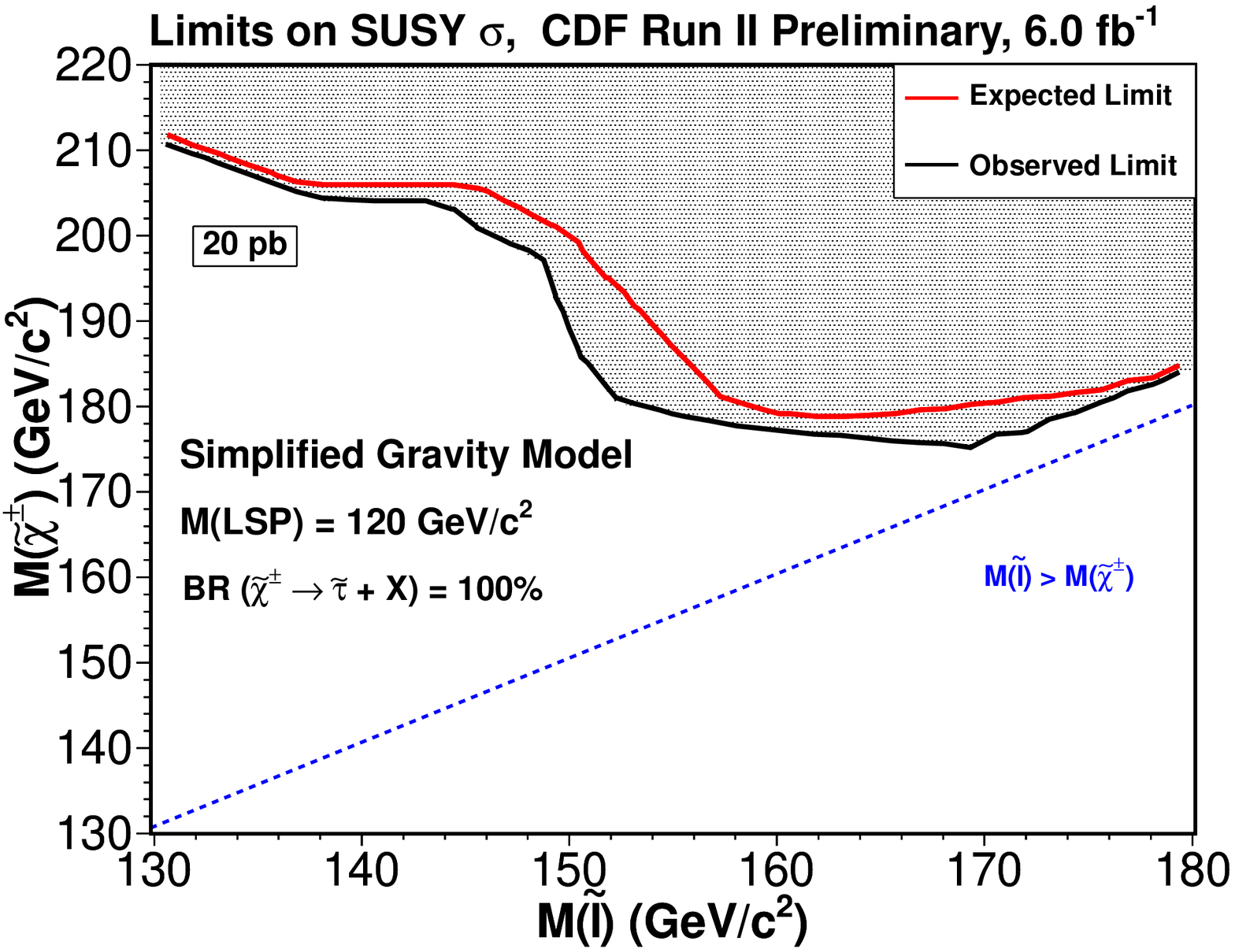} &     \includegraphics[width=8cm,clip=]{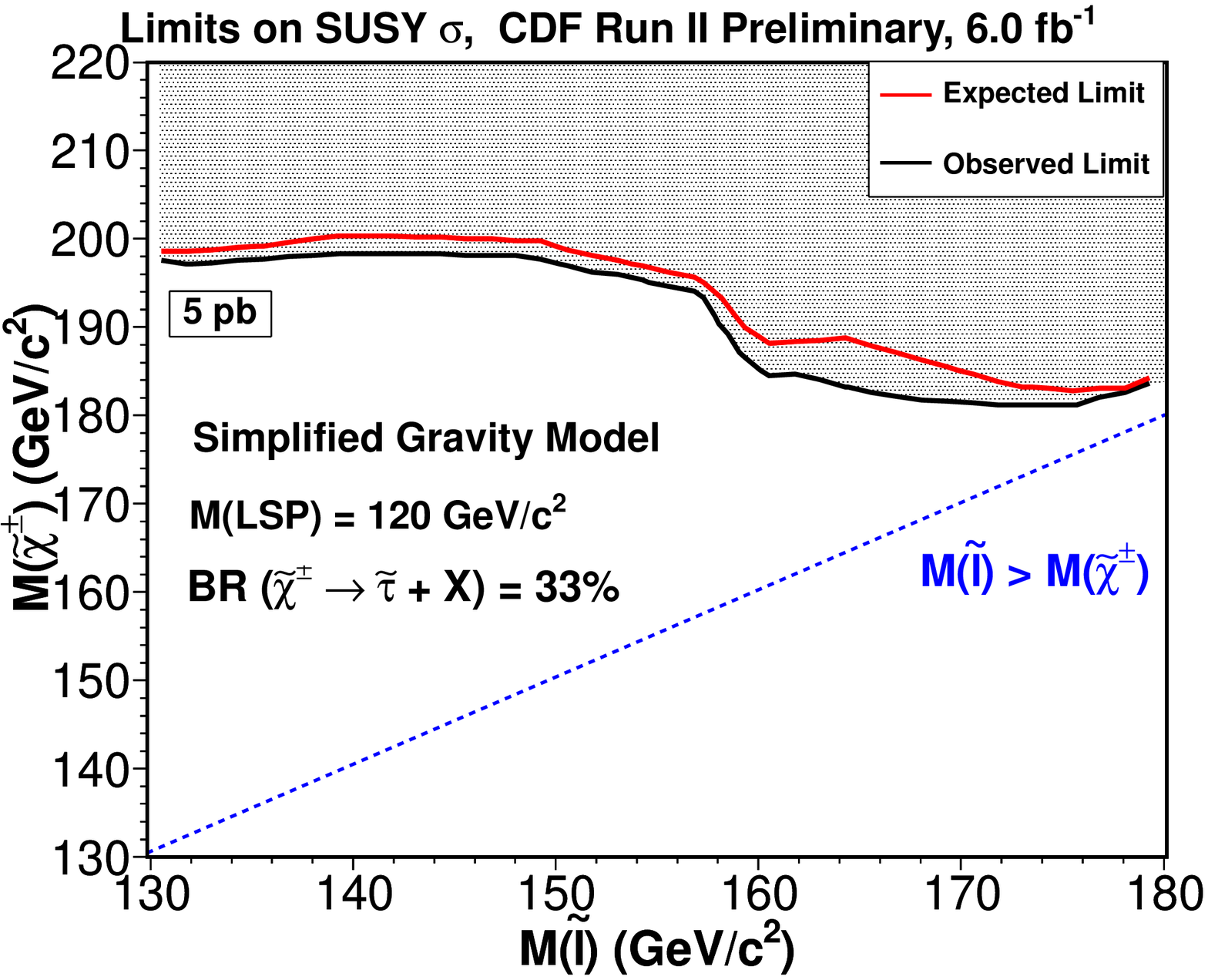}   \\
  \end{tabular}
    \caption{Expected limits (pb) for Simplified Gravity Model with LSP = 120 GeV for BR to taus of 100\% (left), 33\% (right).  \label{fig:exp_gravity_lsp120_2d}}
\end{figure}

\begin{figure}[h!]
  \begin{tabular}{cc}
    \includegraphics[width=8cm,clip=]{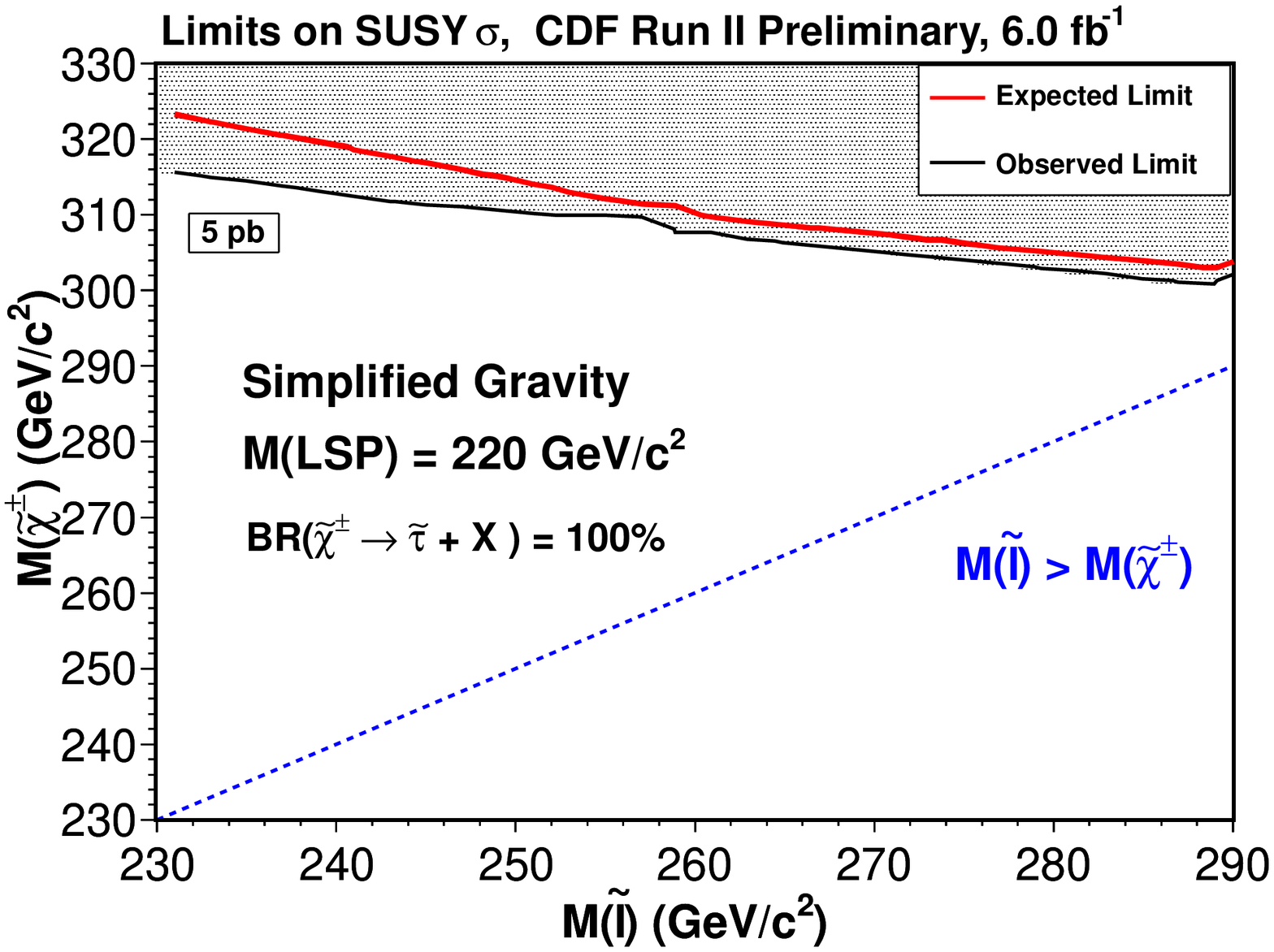} &    \includegraphics[width=8cm,clip=]{figs/limits/gravity_lsp220_c0.eps}  \\
  \end{tabular}
    \caption{Expected limits (pb) for Simplified Gravity Model with LSP = 220 GeV for BR to taus of 100\% (left), 33\% (right).  \label{fig:exp_gravity_lsp220_2d}}
\end{figure}

After the $\met$ cut is applied at each point, we find SUSY production cross section limits and interpolate these contours in the M(Chargino) vs. M(Slepton) plane. The final results can be found in Figures \ref{fig:exp_gauge_2d} through \ref{fig:exp_gravity_lsp220_2d}.

\bigskip 
\clearpage

\end{document}